\documentclass[sigconf]{acmart}

\makeatletter
\def\@ACM@checkaffil{% Only warnings <<<<<<<<<<<<<<<<
    \if@ACM@instpresent\else
    \ClassWarningNoLine{\@classname}{No institution present for an affiliation}%
    \fi
    \if@ACM@citypresent\else
    \ClassWarningNoLine{\@classname}{No city present for an affiliation}%
    \fi
    \if@ACM@countrypresent\else
        \ClassWarningNoLine{\@classname}{No country present for an affiliation}%
    \fi
}
\makeatother

\copyrightyear{2021}
\acmYear{2021}
\setcopyright{acmcopyright}\acmConference[CIKM '21]{Proceedings of the 30th ACM International Conference on Information and Knowledge Management}{November 1--5, 2021}{Virtual Event, QLD, Australia}
\acmBooktitle{Proceedings of the 30th ACM International Conference on Information and Knowledge Management (CIKM '21), November 1--5, 2021, Virtual Event, QLD, Australia}
\acmPrice{15.00}
\acmDOI{10.1145/3459637.3482201}
\acmISBN{978-1-4503-8446-9/21/11}

\newcommand\figref[1]{{Figure \ref{fig:#1}}}
\newcommand\tabref[1]{{Table \ref{tab:#1}}}
\AtBeginDocument{%
  \providecommand\BibTeX{{%
    \normalfont B\kern-0.5em{\scshape i\kern-0.25em b}\kern-0.8em\TeX}}}

\begin{document}
\fancyhead{}

\title{The Effect of News Article Quality on Ad Consumption}

\author{Kojiro Iizuka}
\affiliation{%
  \institution{Gunosy Inc. / University of Tsukuba}
}
\email{iizuka.kojiro@gmail.com}

\author{Yoshifumi Seki}
\affiliation{%
  \institution{Gunosy Inc.}
}
\email{yoshifumi.seki@gunosy.com}

\author{Makoto P. Kato}
\affiliation{%
  \institution{University of Tsukuba / JST, PRESTO}
}
\email{mpkato@acm.org}

\renewcommand{\shortauthors}{Trovato and Tobin, et al.}
\begin{abstract}
  Practical news feed platforms generate a hybrid list of news articles and advertising items (e.g., products, services, or information) and many platforms optimize the position of news articles and advertisements independently.
However, they should be arranged with careful consideration of each other, as we show in this study,
since user behaviors toward advertisements are significantly affected by the news articles.
This paper investigates the effect of news articles on users' ad consumption and shows the dependency between news and ad effectiveness.
We conducted a service log analysis and showed that sessions with high-quality news article exposure had more ad consumption than those with low-quality news article exposure.
Based on this result, we hypothesized that exposure to high-quality articles will lead to a high ad consumption rate.
Thus, we conducted million-scale A/B testing to investigate the effect of high-quality articles on ad consumption,
in which we prioritized high-quality articles in the ranking for the treatment group.
The A/B test showed that the treatment group's ad consumption, such as the number of clicks, conversions, and sales, increased significantly while the number of article clicks decreased.
We also found that users who prefer a social or economic topic had more ad consumption by stratified analysis.
These insights regarding news articles and advertisements will help optimize news and ad effectiveness in rankings considering their mutual influence.
\end{abstract}



%
% Keywords. The author(s) should pick words that accurately describe
% the work being presented. Separate the keywords with commas.

\maketitle

\section{Introduction}
\label{sec:introduction}

News recommendation systems have played an important role in satisfying users' information needs,
as a large amount of news items are produced daily on the Web, which prevents users from finding information of their interests.
It is necessary for news providers to present news articles that satisfy users and, at the same time, generate revenue to deliver service value to users sustainably.
Thus, practical news services should provide users with rankings consisting of both news articles and advertisements (e.g., products, services, and information) to satisfy users and generate revenue.

The position of the news articles and ads
has been optimized based on user behaviors such as click behaviors,
with a strong assumption that articles and ads are independent from each other when applied to real world services~\cite{jointly, allocation}.
However, as we show in this work, user behaviors toward ads are affected by news articles, and, accordingly, 
they need to be arranged with careful consideration of their mutual influence. 

In this paper, we investigate the effect of news articles on ads and show that the ad effectiveness depends significantly on the news quality.
We started with a service log analysis of ad effectiveness with different articles placed together.
The results showed that users exposed to the high-quality article consumed more ads than those exposed to the low-quality article.
We then hypothesized that the exposure to high-quality articles leads to a high ad consumption rate.
Motivated by this hypothesis, we conducted million-scale A/B testing to answer the following research questions:
\begin{itemize}
    \item {\bf RQ1:} To what extent does the quality of news articles affect ad consumption?
    \item {\bf RQ2:} What kinds of users are likely to be affected by the news article quality
    in terms of ad consumption?
    \item {\bf RQ3:} What kinds of ads are likely to be affected by the news article quality
    in terms of their consumption?
\end{itemize}
In the treatment group in A/B testing, we prioritized high-quality articles in the ranking while we did not explicitly prioritize high-quality articles in the control group.
Our A/B testing revealed the following findings:
\begin{itemize}
    \item {\bf RQ1:} The number of clicks, conversions, and sales increased significantly in the treatment group.
    \item {\bf RQ2:} Ad consumption increased regardless of user action volume. Notably, ad consumption of information seekers increased, and there was a significant interaction between news quality and the behavioral intention to use the service.
    \item {\bf RQ3:} The ad consumption increased for most ad genres. The ad consumption of financial ads increased significantly.
\end{itemize}
These insights into the relationship between news articles and ads will help to simultaneously optimize both news and ad effectiveness in rankings.

\section{Related Work}
\label{sec:relatedwork}
This section reviews studies on two research topics related to the current work: user behavior analysis and optimization of organic content and ads. 
% Note that, for comprehensive discussion, {\it documents} were used to refer to both search results and news articles in this section.

\subsection{User Behavior Analysis}
In the field of information retrieval and recommendation systems, users' click behaviors have been studied for a long time~\cite{dupret2008user, hu2008collaborative, pan2008one}.
These studies were conducted from different view points, including content position~\cite{joachims2017accurately, 2007evaluating}, reliability~\cite{yue2010beyond}, quality~\cite{yue2010beyond}, presentation~\cite{wang2013incorporating}, and the delivery mechanism of the system~\cite{lin2018update}.
In this study, we investigate users' click behaviors and ad consumption from the viewpoint of the quality of news articles.

There are reports that the quality of contents in the web site (e.g., news articles and ads) affects user behaviors.
Lu et al. showed that low-quality articles are more likely to be clicked~\cite{quality}.
Lu et al. also investigated the effects of negative user experience in mobile news services and showed that negative experiences reduced user satisfaction~\cite{negative}.
Mao et al. showed that the usefulness of search results correlates with user satisfaction~\cite{mao2016does}.
Alanazi et al. investigated the impact of ad quality on mobile search engine result pages (SERPs) and showed that users pay different amounts of attention to organic results and SERPs for different levels of ad quality~\cite{alanazi2020impact}.
Cramer reported that high-quality native ads could still have a negative effect on perceived site quality if they were too content-relevant ~\cite{cramer2015effects}.

While the effect of the content quality on content own consumption has been investigated, 
the effect of news quality on ad consumption has not yet been studied in the literature, and is first revealed in this work.
% Based on a large scale user experiment, we studied the effects of news article quality on ad consumption.

\subsection{Optimization of Organic Content and Advertisements}

Web services, such as social network services and news services, provide users with organic content and advertisements.
Organic content attracts users, 
while advertisements are a source of revenue for many companies. 
Both are vital for sustaining services.
Therefore, there are many studies focusing on the optimization of user satisfaction from organic content and that of revenue from advertising.
Zhao et al. proposed a method to generate a ranking that considers long-term user experience from organic content and profits from advertisements~\cite{jointly}.
Wang et al. proposed a reinforcement-learning-based method that dynamically inserts advertisements rather than fixing the position and number of advertisements to be inserted~\cite{adaptive}.
Yan et al. proposed a method for optimizing the location of ads inserted between articles in a news feed~\cite{allocation}.

These studies treated organic content and advertisements as independent from each other when applied to real-world services.
In this study, we show that the quality of news articles affects ad consumption.
If we built our insight into the optimization problem, the optimization would be performed more effectively.

\section{Preliminary}
\label{sec:preliminary}
In this section, we describe a news service that we used for our study 
and define the quality of news articles.
% In addition, we introduce some metrics used in the analysis.

\subsection{Target Service}
\label{targetservice}
% We used one of a most popular mobile news services Gunosy for our study\footnote{The service name and country are anonymized for double-blind review.}.
We analyzed user behavior on Gunosy, \footnote{https://gunosy.com} one of the most popular news applications in Japan. %%mod
When a user visits the news service, he or she is given a ranking including news articles based on the his or her interest.
At the same time, in-feed ads are presented in between news articles.
When a user clicks on the title or image of an article in the ranking, 
the system displays the detailed content of the clicked article.
Users can stop reading the content at any time, and,
if stopped, they can return to the clicked position in the original ranking.
% Reading of the contents can be interrupted at any time, and if interrupted, the system returns to the clicked position in the original ranking.
In this way, a user can browse multiple articles in a ranking.
The user can quit at any time during these series of behaviors.

The news platform we used collects user feedback including the scroll depth in an article, ad conversions\footnote{{\it Conversion} refers to the user behavior of purchasing products, installing other services introduced in the advertisement, or requesting detailed information about the advertised product.}, and clicks and impressions for both the articles and ads.
Articles and ads are displayed to users in a ranking, 
and data are collected for each position in the ranking.

\subsection{Article Quality}

\subsubsection{Definition}
Based on an existing study~\cite{quality}, 
article quality is defined as a degree correlating to the following criteria.

\begin{description}
\item[{\bf Authenticity}] The content is authentic, has high credibility, and is not imaginary or exaggerated~\cite{vosoughi2018spread}.
\item[{\bf Expression}] The content is objective, accurate, and the information volume is adequate but not redundant.
In addition, the content is not vulgar, violent, or bloody, and does not contain pornographic words.
\item[{\bf Headline}] The headline information is consistent with the body content and is not fake, exaggerated, or vulgar.
\end{description}
% Considering these quality criteria, we classify articles into high quality, low quality, and others.
In this paper, we define high-quality articles as those satisfying all of the criteria, 
whereas low-quality articles are those violating at least one of the criteria.
We note that our definition of low-quality articles is related to the concept of {\it clickbait}~\cite{blom2015click, chen2015misleading}.

\subsubsection{Quality Judgement}
We judged the quality of news articles by using experts and a rule-based method.
News articles to be judged were selected from six popular news categories: 
social, economic, entertainment, technology, sport, and cooking.
First, we developed a list of low-quality words, such as vulgar, violent, or cruel words.
Then, we retrieved articles containing one or more low-quality words in their headline or body,
and judged them as low-quality.
Experts were employed to find high-quality articles because it is difficult to judge the objectivity and accuracy of the content and appropriate length of articles by simply using heuristics~\cite{quality}.
Three experts were involved in identifying a high-quality article.
An article is considered high-quality if it is judged by all experts to meet the quality criteria.

These experts were members of the content quality team working for the news platform.
The content quality team is comprised of professionals 
involved in news recommendations from a qualitative perspective.
For example, they are responsible for preventing the diffusion of news articles that could be harmful.

% \subsection{Metrics}
% We used the following metrics to measure the consumption of news articles and ads:
% \begin{description}
% \item[{\bf Click-Through Rate (CTR)}] This metric is defined as the number of clicks divided by the number of impressions. CTR is used for both article and ad analysis. 
% CTR is a metric of the likelihood that users will click on the article or ad.
% \item[{\bf Article Scroll Rate}]  This metric is defined as 
% the length of the article read by the user divided by the total length of the article. 
% If this metric is low, it may indicate that users are not likely to have read the article until the end.
% \item[{\bf Conversion Rate (CVR)}] This metric is defined as the number of conversions divided by the number of clicks. When this metric is low, it suggests that many users clicked on the ad but did not purchase it.
% \item[{\bf Sales/Users}] This metric is defined as the sales divided by the number of unique users.
% The sales come from advertisements when a user clicks.
% \end{description}
% Since there is a business risk to exposing raw metrics, 
% all values are normalized to $[0, 1]$, 
% but their relative differences are shown in the \figref{logarticleconsumption}, \figref{logadconsumption} and \figref{segment}.

\section{Log Analysis}
\label{sec:logbasedanalysis}
\begin{figure}[tb]
\begin{tabular}{cc}

\begin{minipage}{0.5\hsize}
\begin{center}
\includegraphics[width=40mm]{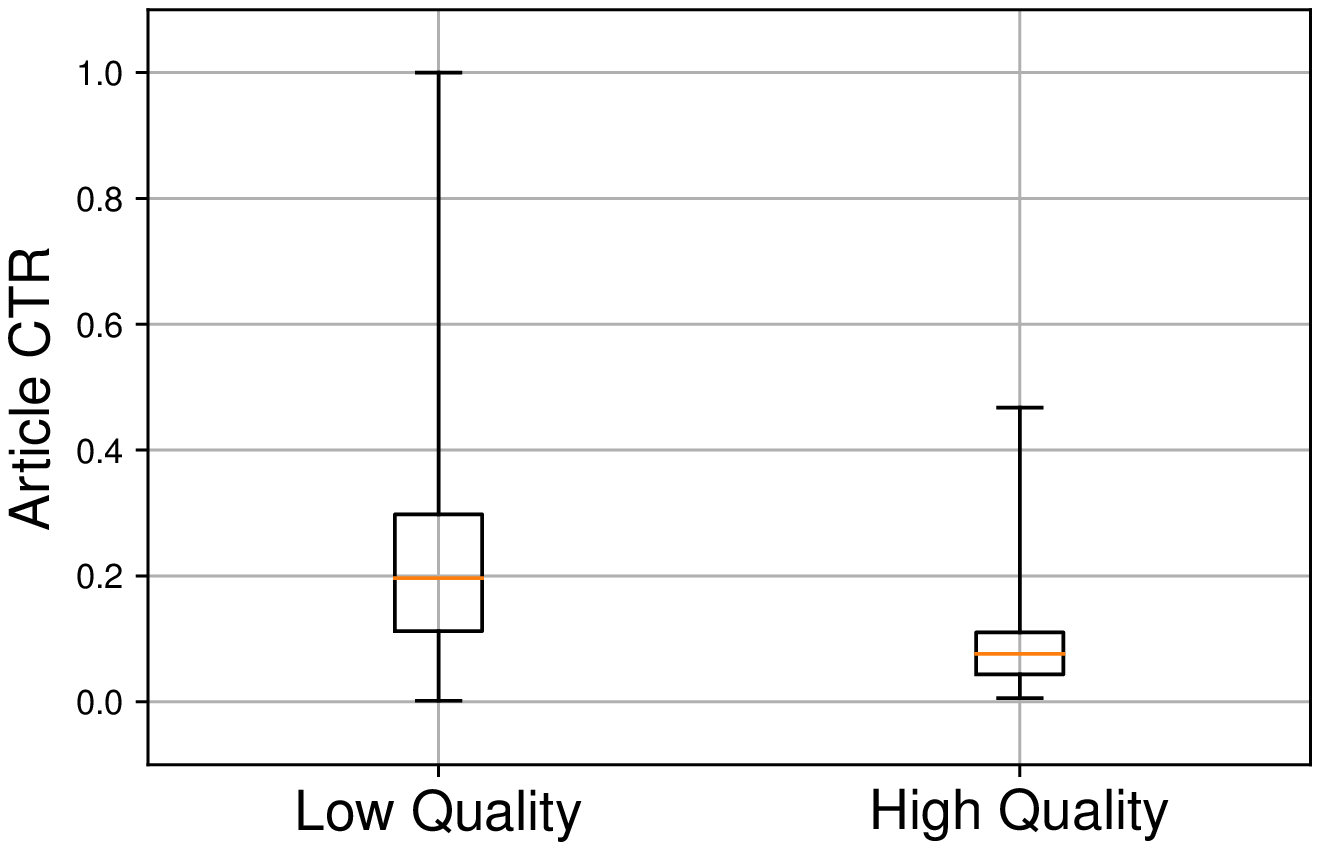}
\end{center}
\end{minipage}

\begin{minipage}{0.5\hsize}
\begin{center}
\includegraphics[width=40mm]{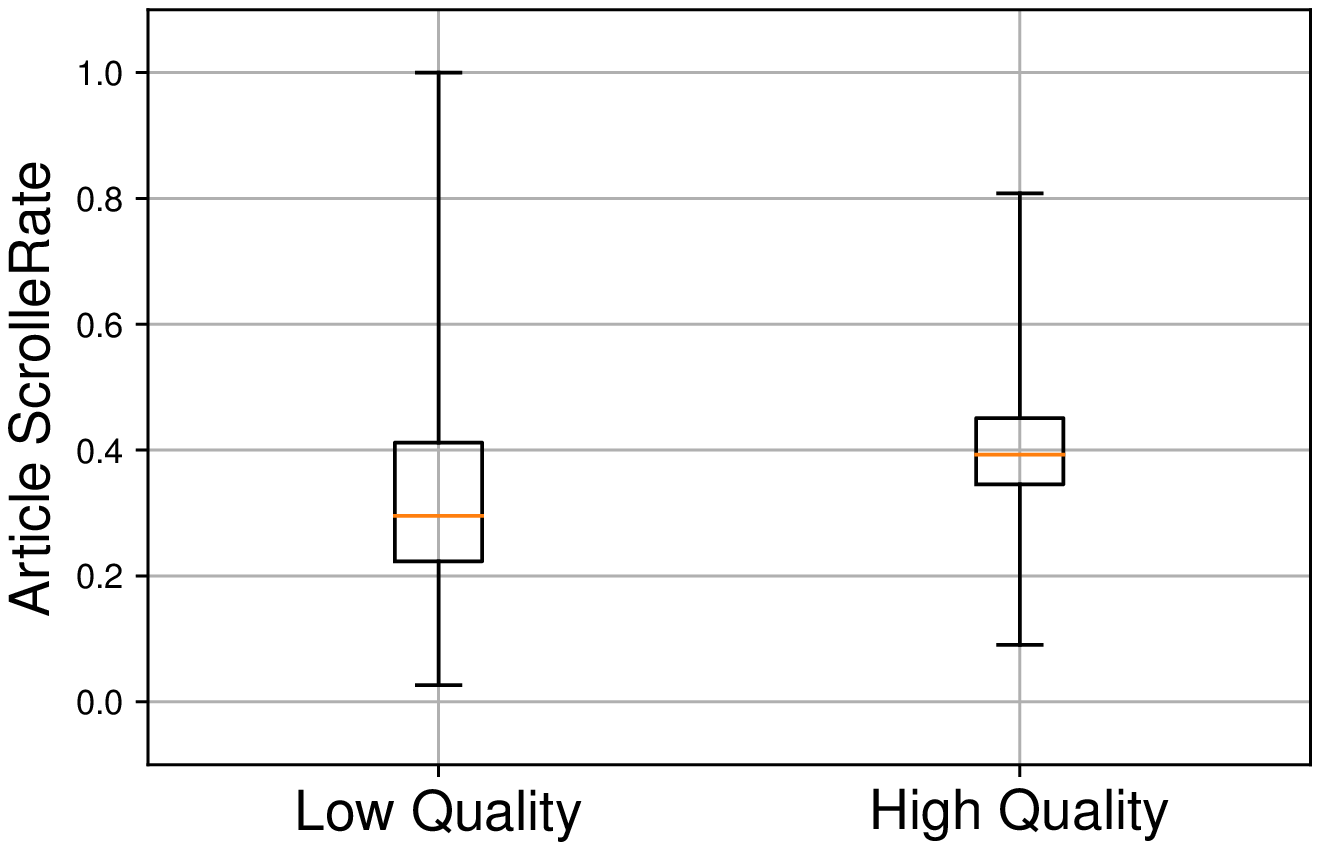}
\end{center}
\end{minipage}

\end{tabular}
\caption{Quality Effect on Article Consumption.}
\label{fig:logarticleconsumption}
\end{figure}

In this section, we investigate the effect of of news article quality on ad consumption through service log analysis.\footnote{Since there is a business risk to exposing raw metrics, all values are normalized to $[0, 1]$, but their relative differences are shown in the \figref{logarticleconsumption}, \figref{logadconsumption} and \figref{segment}.}

% \begin{figure}[tb]
% \begin{tabular}{cc}
% \begin{minipage}{0.5\hsize}
% \begin{center}
% \includegraphics[width=35mm]{images/High-qualityCategoryDistribution-uplift.eps}
% \end{center}
% \end{minipage}

% \begin{minipage}{0.5\hsize}
% \begin{center}
% \includegraphics[width=35mm]{images/Low-qualityCategoryDistribution-uplift.eps}
% \end{center}
% \end{minipage}

% \end{tabular}
% \caption{Category Distributions.}
% \label{fig:categorydistribution}
% \end{figure}

\subsection{Data}
We collected 30 days of user feedback on a million-scale news service described in Section~\ref{targetservice}.
During this period, the number of high-quality articles was 537 and the number of low-quality articles was 4,472.
The total number of user sessions\footnote{Session means the time interval between when a user opens and closes the mobile news service.} in the collected logs was more than 10 million.
To reduce bias, such as different amounts of user activity and position bias, 
we only analyzed the user behaviors at {\it top section} of the ranking,
which consists of three articles at the first, second and third position and one advertisement at the fourth position in the ranking.

\begin{comment}
\figref{categorydistribution} shows the distribution of news categories according to annotated article quality used in the log analysis.
High-quality articles are mainly in the social category, which includes articles on politics and economics.
On the other hand, entertainment articles account for almost all of the low-quality articles.
\end{comment}

\subsection{\bf Article Consumption}
% \subsubsection{result}
We first investigated the effect of article quality on the article consumption.
\figref{logarticleconsumption} shows the relationship between the article quality 
and article consumption measured by the article CTR and scroll rate.
The figure is a box-and-whisker plot of the article CTR and scroll rate grouped by the article quality.
The median article CTR for low-quality articles is higher than the median article CTR for high-quality articles ($p < 0.05$, chi-square test for the average value), and the median article scroll rate for low-quality articles is lower than that for high-quality articles ($p < 0.05$, unpaired t-test for the average value).

These results suggest that low-quality articles are more likely to be clicked, 
but not be read to the end compared to high-quality articles.
% In this aspect, reactions to low-quality articles are similar to those to clickbait articles.
% This similarity might be because our definition of low-quality articles is comparable to that of clickbait.
It is consistent with previous studies~\cite{quality} that
the result of article CTR and scroll rate vary for different levels of article quality.
% Thus, we confirm that the quality we classify affects the article's action the same as the previous work.

\subsection{\bf Ad Consumption}
\begin{figure}[tb]
\begin{tabular}{cc}
\begin{minipage}{0.5\hsize}
\begin{center}
\includegraphics[width=40mm]{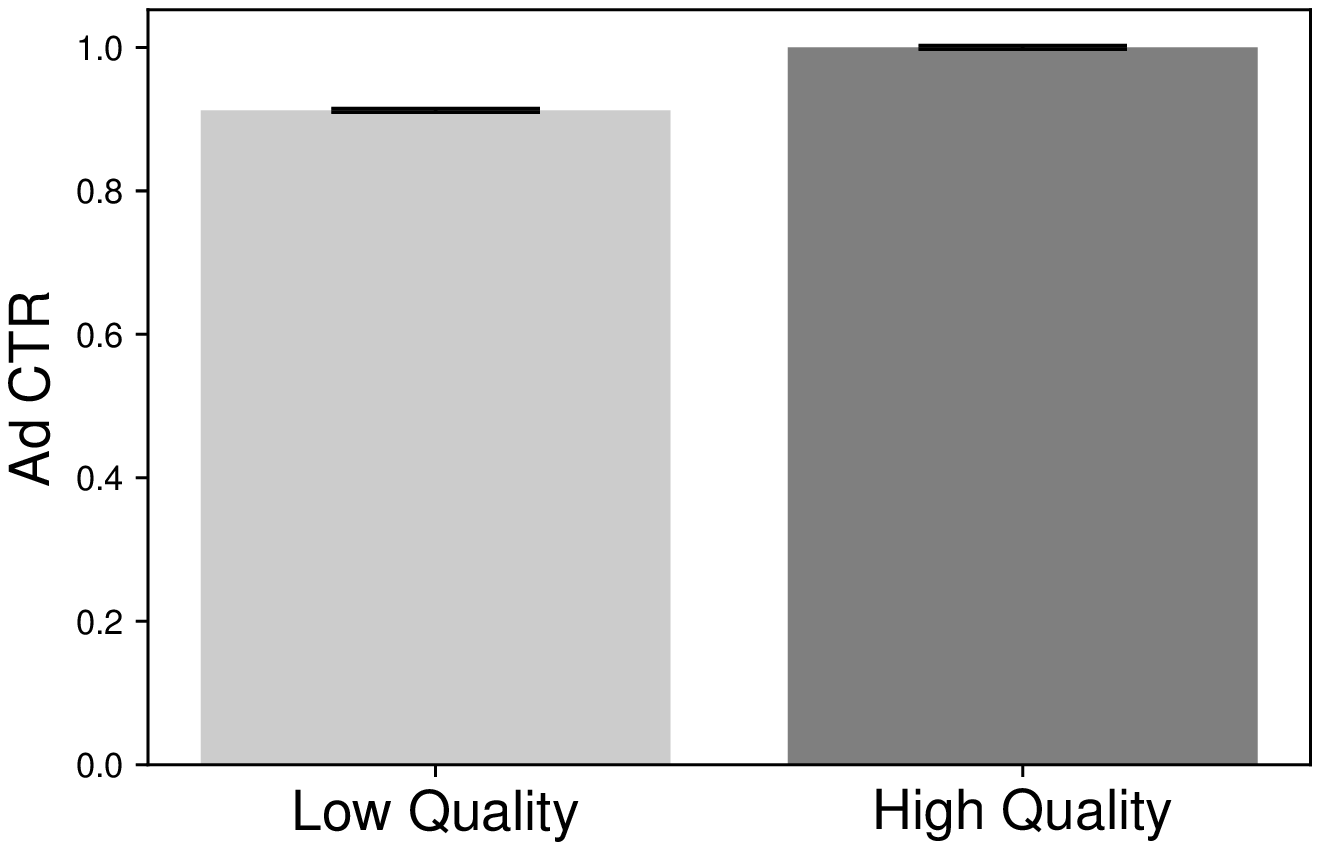}
\end{center}
\end{minipage}

\begin{minipage}{0.5\hsize}
\begin{center}
\includegraphics[width=40mm]{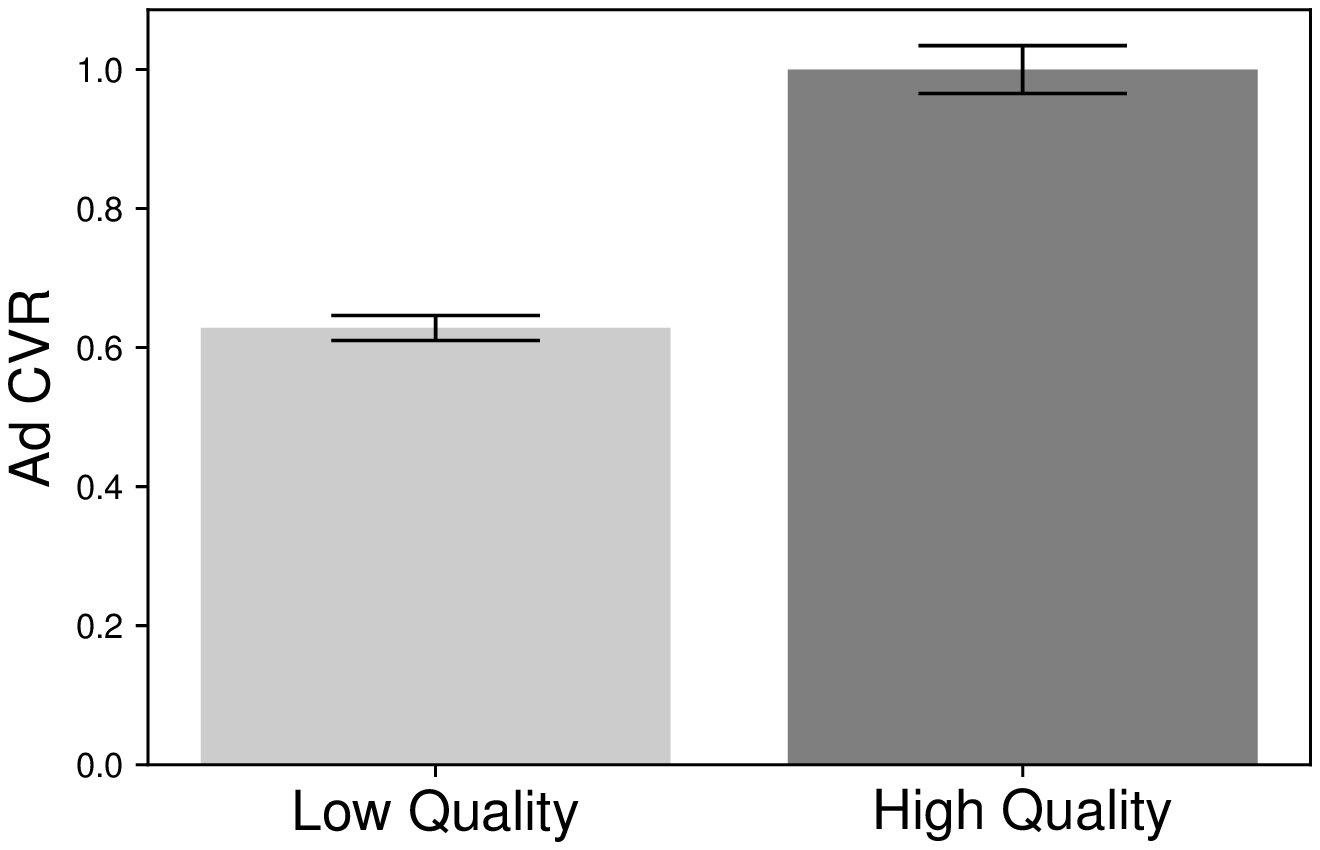}
\end{center}
\end{minipage}

\end{tabular}
\caption{Quality Effect on Ad Consumption. Error bar shows the 95\% confidence interval for the Ad CTR and CVR.}
\label{fig:logadconsumption}
\end{figure}

We then investigate the effect of article quality on the ad consumption.
\figref{logadconsumption} shows the ad CTR and CVR for two conditions
in which the ad was exposed together with a high- or low-quality article in the ranking.
Note that, if both high- and low-quality articles exist in the ranking, 
the denominator of metrics for both high- and low-quality 
(i.e., the number of impressions or clicks) was incremented.
The left figure in \figref{logadconsumption} shows the ad CTR, 
while the right figure shows the ad CVR.
When an advertisement is exposed with high-quality articles, 
the ad CTR and CVR are higher than those with low-quality articles 
($p < 0.05$, chi-square test for each of ad CTR and CVR).

The result of the ad CTR may imply that low-quality articles got more users' attention than the advertisement, which led to the ad CTR being compromised.
Another possible explanation for a high CVR is that users are more likely to trust the service with a high-quality article, which results in deeper engagement in the service  (i.e., conversion).

\section{Online Experiment}
\label{sec:onlineexperiment}
From the results of our log analysis, we hypothesize that displaying high-quality articles can lead to higher ad consumption.
In this section, we conduct million-scale A/B testing to answer the following research questions:

\begin{itemize}
    \item {\bf RQ1:} To what extent does the quality of news articles affect ad consumption?
    \item {\bf RQ2:} What kinds of users are likely to be affected by the news article quality
    in terms of ad consumption?
    \item {\bf RQ3:} What kinds of ads are likely to be affected by the news article quality
    in terms of their consumption?
\end{itemize}

\subsection{Setting}
As in the log analysis, we conducted A/B testing on online news services described in Section~\ref{targetservice}.
The duration of the A/B test was one month, which did not overlap with the log analysis period.
In this A/B testing, we only focused on the effect of high-quality articles.
We were not allowed to expose low-quality articles intentionally,
since it can sacrifice user experience and ad effectiveness 
as the log analysis suggested.

The A/B testing target users in the treatment and control groups were randomly assigned using a 1:1 ratio from a subset of all users.
High-quality articles were preferentially presented for users in the treatment group, more specifically, they were inserted into the {\it top section}, 
or the top three positions in the ranking.
On the other hand, the ranking generated from the existing recommendation algorithm was displayed in the control group.
High-quality articles were those used in the log analysis.
% The timing of selecting high-quality articles was set to at least once every five hours between 7:00 a.m. and 10:00 p.m.
High-quality articles presented to users were rotated 
at least once every five hours between 7:00 a.m. and 10:00 p.m.

\subsection{Results}
\subsubsection{{\bf RQ1:} To what extent does the quality of articles affect advertisement consumption?}

\begin{table}[t]
\begin{center}
\caption{Overall A/B Testing Result. Ad consumption was increased significantly in the treatment group for both iOS and Android users.}
\label{tab:overall}
\begin{tabular}{|r|cccc|}
\hline
& AdCTR & AdCVR & Sales/Users & ArticleCTR\\ \hline
iOS User & +1.71\% & +2.37\% & +3.61\% & -3.19\% \\
Android User & +3.17\% & +4.27\% & +5.59\% & -4.13\% \\ \hline
\end{tabular}
\end{center}
\end{table}

\tabref{overall} shows the overall results of A/B testing.
The results are divided into iOS users and Android users because we know from past experiments that results differ depending on the OS type.

As an overall result, all advertising metrics such as CTR, CVR, and Sales/Users  improved significantly in the treatment group ($p < 0.05$ in chi-square test for AdCTR and AdCVR, and $p < 0.05$ in unpaired t-test for Sales/Users).
This result indicates that high-quality articles increased the consumption of ads.
We note that the article CTR was decreased in the treatment group ($p < 0.05$, chi-square test).
As can also be seen in the log analysis, 
high-quality articles received fewer clicks than low-quality articles.

To answer {\bf RQ1:} AdCTR, AdCVR, and Sales/Users improved significantly in the treatment group in which high-quality articles were presented at the top section of the ranking.

\subsubsection{{\bf RQ2:} What kinds of users are likely to be affected by the news article quality in terms of ad consumption?}

\begin{figure}[tb]
\begin{tabular}{cc}
\begin{minipage}{0.5\hsize}
\begin{center}
\includegraphics[width=40mm]{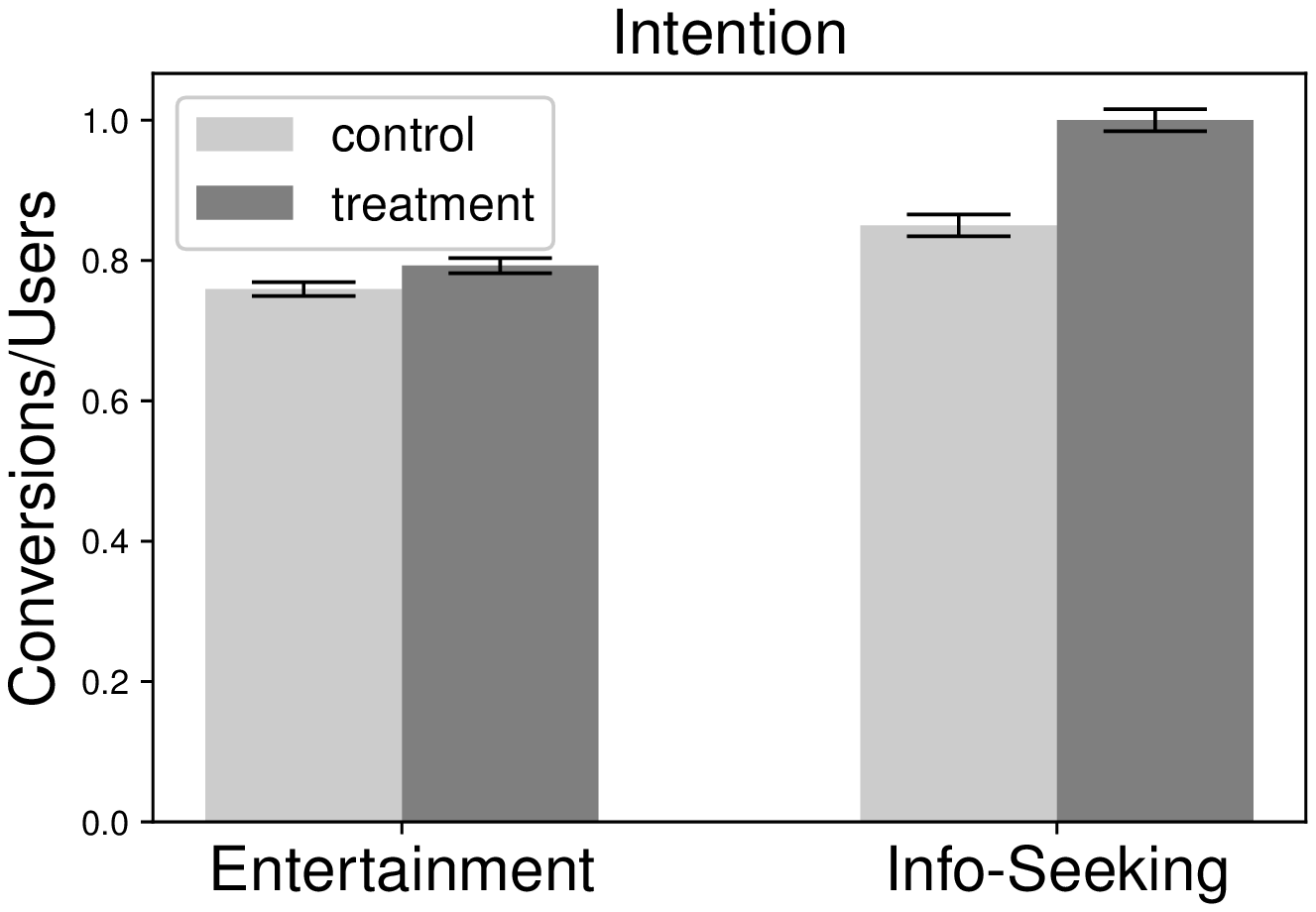}
\end{center}
\end{minipage}

\begin{minipage}{0.5\hsize}
\begin{center}
\includegraphics[width=40mm]{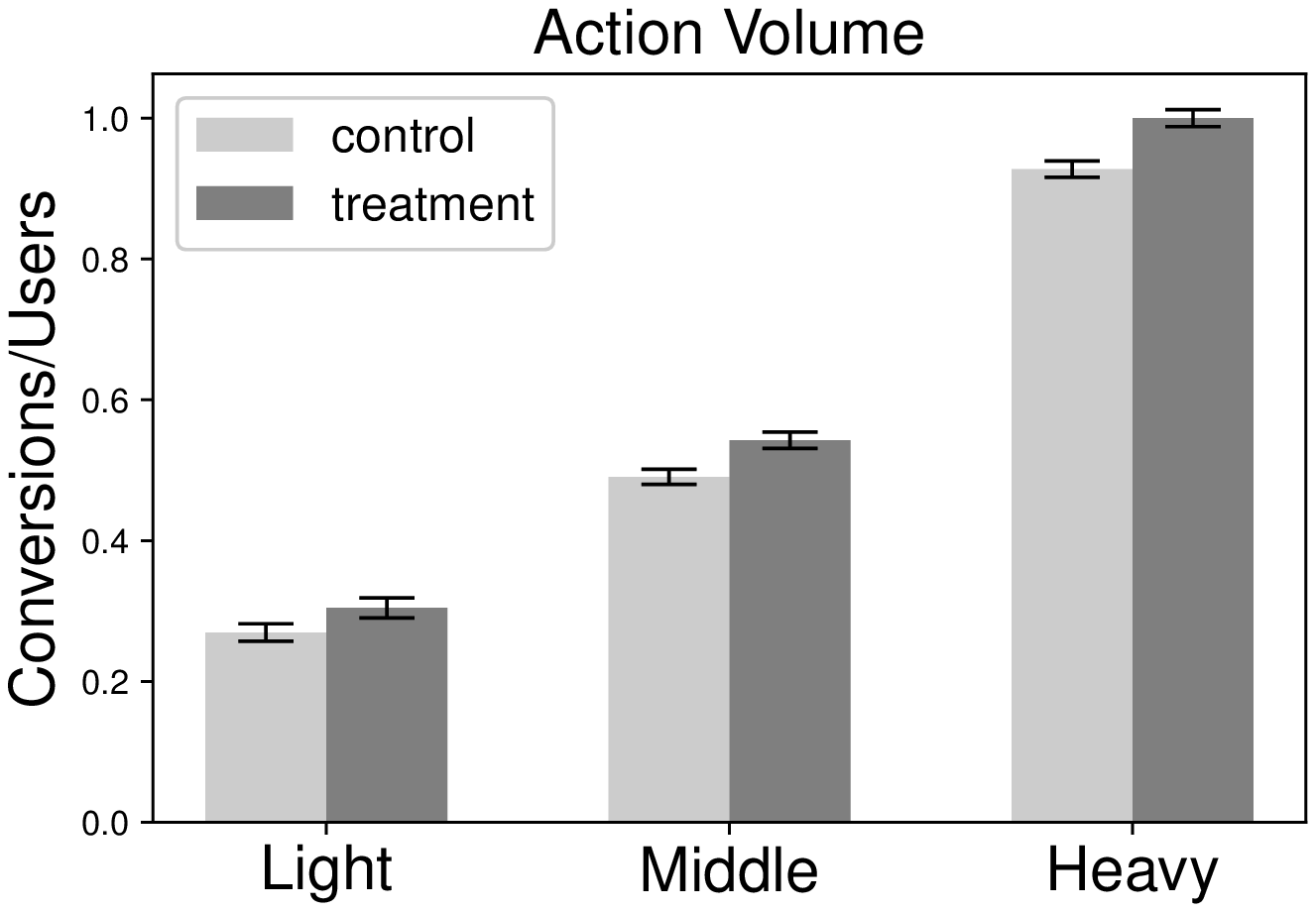}
\end{center}
\end{minipage}

\end{tabular}
\caption{Ad Consumption by User Segment.}
\label{fig:segment}
\end{figure}

To investigate in detail the effects of article quality, 
we compare the results by user segments.
We segmented users by two factors:
intention to use the news service and action volume.

We divided the users into two groups by their intention: 
information-seeking and entertainment.
Inspired by previous work~\cite{amazeen2020news}, we define information-seekers as those whose clicks for economic or social topics account for more than 20\% of their clicks in the past month, and users with entertainment intention as the others.
We set the threshold at 20\% to balance the sample size of the both groups.

Action volume is measured by the number of user clicks in the past month,
and is divided into three groups: light, middle, and heavy.
We used the 33.3\% and 66.6\% percentiles for the number of clicks 
to divide users by the action volume.
Users in the 33\% percentile or lower were considered light users whereas users in the 66\% percentile or above were considered heavy users. Middle users fell in between.

\figref{segment} shows the differences of the number of conversions grouped by the user segments,
in which the left figure shows those by the user intention 
and right figure shows those by the action volume.
We can see that information seekers in the treatment group were more influenced than 
users with entertainment intent in the treatment group.
By action volume, all type of users had more ad consumption in the treatment group.

We conducted regression analysis by using a generalized linear model (GLM)~\cite{nelder1972generalized} with a binomial distribution and a logit link function.
To see the interaction between the A/B testing group and both intention and action volume, we formulated the regression model to predict conversion as follows: conversion $\sim$ group $+$ intention $+$ action volume $+$ group $\times$ intention + group $\times$ action volume.
The regression result shows that the interaction term of the A/B testing group and the intention was significant ($p < 0.05$, Wald test) while the interaction term of the A/B testing group and the action volume was not significant.

Interestingly, information seekers were more likely to accept ads.
This may imply that information seekers had more attention 
toward ads because of the high-quality articles placed above the ads, 
which contained objective and accurate information.

To answer {\bf RQ2:} Ad consumption increased regardless of the user intention or user action volume.
Especially, ad consumption of information seekers  increased, and there existed an interaction between the news quality and user intention.

\subsubsection{{\bf RQ3:} What kinds of ads are likely to be affected by the news article quality in terms of their consumption?}
\begin{figure}[t]
\includegraphics[scale=0.4]{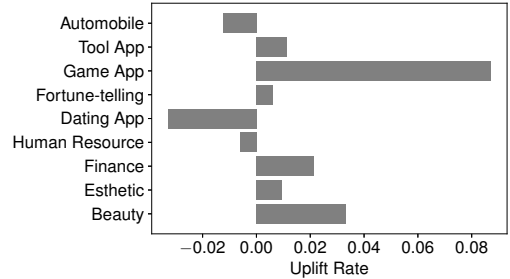} %Genres-tr-and-ct
\caption{Ad Consumption by Ad Genre.}
\label{fig:genre}
\end{figure}
\figref{genre} shows the uplift rate of conversions per ad genre.
We can see that the ad consumption was boosted by high-quality articles for
most of the ad genres except for {\it Dating App} and {\it Automobile}.
A chi-square test with Holm correction revealed that the {\it Finance} genre was the only genre that was statistically significant ($p < 0.05$), and this was beneficial to our news platform.
Because the ad consumption for Finance genre was less than the other news services in the country
where the news platform is located, 
increasing the ad consumption for the Finance genre was challenging for our news platform.
% It was surprising to us that ad consumption in certain genres was affected by the quality of news articles.

To answer {\bf RQ3:} 
Most of the genres showed increase in ad consumption, 
among which the increase for the finance genre was significant.

% The results of these experiments lead to the decision to build article quality into news recommendation algorithms to optimize both article and ad metrics.

\section{Conclusion}
\label{sec:conclusion}
In this study, we investigated the impact of article quality on  ad consumption in news services.
To investigate the effect, we first conducted a log-based  analysis.
Ad consumption was better for high-quality articles compared to low-quality articles.
To verify in detail, we conducted a million-scale online experiment using A/B testing.
The A/B test results showed that ad CTR, CVR, and sales improved significantly in the treatment group that explicitly involved high-quality articles.
We also found that the ad consumption of users who prefer social or economic topics increased.
These insights regarding news and advertisements will help optimize news and ad effectiveness in rankings.

% Future work should include a more detailed investigation.% of the effect of the quality of news articles.
In the future work, we plan to conduct a study to investigate the effect of the user's subjective perception of news on their response to an advertisement.

\bibliographystyle{ACM-Reference-Format}
\balance
\bibliography{main}

\end{document}